\documentclass[12pt,a4paper,twoside]{article}
\RequirePackage{latexsym,amsmath,amssymb,amsfonts,amsthm}
\RequirePackage[dvipsnames,usenames]{color}
\pdfoutput=1
\usepackage{a4wide}
\usepackage{amssymb}
\usepackage{bm}
\usepackage{dcolumn}
\usepackage{amsfonts}
\usepackage{graphicx,epsfig}
\usepackage{psfrag}
\usepackage{multicol}
\usepackage{tikz}
\usetikzlibrary{arrows,shapes}
\usetikzlibrary{matrix,arrows}
\newcommand{\be}{\begin{eqnarray}}
\newcommand{\ee}{\end{eqnarray}}
\newcommand{\bra}[1]{\mbox{$\langle\, #1 \mid$}}

\newcommand{\ket}[1]{\mbox{$\mid #1\,\rangle$}}

\newcommand{\pro}[2]{\mbox{$\langle\, #1 \mid #2\,\rangle$}}
\newcommand{\expec}[1]{\mbox{$\langle\, #1\,\rangle$}}

\renewcommand{\d}{\mbox{${\rm d}$}} 
\newcommand{\lp}{\ell_{\rm p}}
\newcommand{\mpl}{m_{\rm p}}
\newcommand{\gn}{G_{\rm N}}
\newcommand{\rh}{r_{\rm H}}
\newcommand{\Rh}{R_{\rm H}}
\newcommand{\psis}{{\psi}_{\rm S}}

\newcommand{\psih}{{\psi}_{\rm H}}
\newcommand{\en}{{\mathcal{E}}}
\theoremstyle{plain}
\newtheorem{thm}{Theorem}[section]
\newtheorem{corollary}[thm]{Corollary}
\title{\bf Horizon Quantum Mechanics of Rotating Black Holes}
\author{Roberto~Casadio$^{ab}$\thanks{E-mail: casadio@bo.infn.it},
Andrea~Giugno$^{c}$\thanks{E-mail: A.Giugno@physik.uni-muenchen.de},
Andrea~Giusti$^{abc}$\thanks{E-mail: andrea.giusti@bo.infn.it}
$\,$
and
Octavian~Micu$^d$\thanks{E-mail: octavian.micu@spacescience.ro}
\\
\\
$^a${\em Dipartimento di Fisica e Astronomia, Universit\`a di Bologna}
\\
{\em via Irnerio~46, I-40126 Bologna, Italy}
\\
\\
$^b${\em I.N.F.N., Sezione di Bologna, I.S.~FLAG}
\\
{\em via B.~Pichat~6/2, I-40127 Bologna, Italy}
\\
\\
$^c${\em Arnold Sommerfeld Center, Ludwig-Maximilians-Universit\"at}
\\
{\em Theresienstra{\ss}e 37, 80333 M\"unchen, Germany}
\\
\\
$^d${\em Institute of Space Science, Bucharest}
\\
{\em P.O.~Box MG-23, RO-077125 Bucharest-Magurele, Romania}
}
\begin{document}
\maketitle
\begin{abstract}
The Horizon Quantum Mechanics is an approach that was previously introduced in order
to analyse the gravitational radius of spherically symmetric systems and compute the probability
that a given quantum state is a black hole.
In this work, we first extend the formalism to general space-times with asymptotic (ADM)
mass and angular momentum.
We then apply the extended Horizon Quantum Mechanics to a harmonic model of
rotating corpuscular black holes.
We find that simple configurations of this model naturally suppress the appearance of
the inner horizon and seem to disfavour extremal (macroscopic) geometries.
\end{abstract}
\section{Introduction}
\label{intro}
\setcounter{equation}{0}
Astrophysical compact objects are known to be usually rotating, and one correspondingly
expects most black holes formed by the gravitational collapse of such sources be of the Kerr type.
The formalism dubbed Horizon Quantum Mechanics
(HQM)~\cite{fuzzyh,hqft,gupf,acmo,Casadio:2015rwa,Casadio:2015qaq},
was initially proposed with the purpose of describing the gravitational radius of spherically
symmetric compact sources and determining the existence of a horizon in a quantum mechanical fashion.
It therefore appears as a natural continuation in this research direction to extend the HQM to rotating
sources.
Unfortunately, this is not at all a conceptually trivial task.
\par
In a classical spherically symmetric system, the gravitational radius is uniquely defined
in terms of the (quasi-)local Misner-Sharp mass and it uniquely determines the location of the
trapping surfaces where the null geodesic expansion vanishes.
The latter surfaces are proper horizons in a time-independent configuration, which is
the case we shall always consider here.
It is therefore rather straightforward to uplift this description of the causal structure 
of space-time to the quantum level by simply imposing the relation between the
gravitational radius and the Misner-Sharp mass as an operatorial constraint to be satisfied
by the physical states of the system~\cite{hqft}.
\par
In a non-spherical space-time, such as the one generated by an axially-symmetric rotating source,
although there are candidates for the quasi-local mass function that should replace the Misner-Sharp
mass~\cite{qlm-review}, the locations of trapping surfaces, and horizons, remain to be
determined separately.
We shall therefore consider a different path and simply uplift to a quantum
condition the classical relation of the two horizon radii with the mass
and angular momentum of the source obtained from the Kerr metric.
This extended HQM is clearly more heuristic than the one employed for the 
spherically symmetric systems, but we note that it is indeed fully consistent
with the expected asymptotic structure of axially symmetric space-times.
\par
Beside the formal developments, we shall also apply the extended HQM 
to specific states with non-vanishing angular momentum of the harmonic black hole model
introduced in Ref.~\cite{QHBH}~\footnote{See also Ref.~\cite{muck} for an improved version.}.
This model can be considered as a working realisation of the corpuscular black holes
proposed by Dvali and Gomez~\cite{dvali}, and turns out to be simple enough, so as to allow
one to determine explicitly the probability that the chosen states are indeed black holes.
Furthermore, we will investigate the existence of the inner horizon and likelihood of
extremal configurations for these states.
\par
The paper is organised as follows:
at the beginning of Section~\ref{defH}, we briefly summarise the HQM and recall some of the main
results obtained for static spherically symmetric sources;
the extension of the existing formalism to the case of stationary axisymmetric sources, which are both localised
in space and subject to a motion of pure rotation, is presented in Section~\ref{rotS};
a short survey of the harmonic model for corpuscular black holes is given in Section~\ref{BECsection},
where we then discuss some elementary applications of the HQM to rotating black holes whose quantum
state contains a large number of (toy) gravitons;
finally, in Section~\ref{conc}, we conclude with remarks and hints for future research. 
\section{Horizon quantum mechanics}
\label{defH}
\setcounter{equation}{0}
We start from reviewing the basics of the (global) HQM for static spherically symmetric
sources~\cite{fuzzyh,hqft,gupf,acmo,Casadio:2015rwa,Casadio:2015qaq}, and then extend this formalism
to rotating systems by means of the Kerr relation for the horizon radii in terms of the asymptotic mass and
angular momentum of the space-time.
In particular, we shall rely on the results for the ``global'' case of Ref.~\cite{hqft}  and follow closely the
notation therein.
\subsection{Spherically symmetric systems}
\label{sphS}
The general spherically symmetric metric $g_{\mu\nu}$ can be written
as~\footnote{We shall use units with $c=1$,
and the Newton constant $G=\lp/\mpl$, where $\lp$ and $\mpl$
are the Planck length and mass, respectively, and $\hbar=\lp\,\mpl$.}
\be
\d s^2
=
g_{ij}\,\d x^i\,\d x^j
+
r^2(x^i)\left(\d\theta^2+\sin^2\theta\,\d\phi^2\right)
\ ,
\label{metric}
\ee
where $r$ is the areal coordinate and $x^i=(x^1,x^2)$ are coordinates
on surfaces of constant angles $\theta$ and $\phi$.
The location of a trapping surface is then determined by the equation
\be
g^{ij}\,\nabla_i r\,\nabla_j r
=
0
\ ,
\label{th}
\ee
where $\nabla_i r$ is perpendicular to surfaces of constant area
$\mathcal{A}=4\,\pi\,r^2$.
If we set $x^1=t$ and $x^2=r$, and denote the static matter density as
$\rho=\rho(r)$, Einstein field equations tell us that
\be
g^{rr}
=
1-\frac{2\,\lp\,(m/\mpl)}{r}
\ ,
\label{einstein}
\ee
where the Misner-Sharp mass is given by 
\be
m(r)
=
4\,\pi\int_0^r \rho(\bar r)\,\bar r^2\,\d \bar r
\ ,
\label{M}
\ee
as if the space inside the sphere were flat.
A trapping surface then exists if there are values of $r$ such that
the gravitational radius $\rh = 2\,\lp\,{m}/{\mpl}\ge r$.
If this relation holds in the vacuum outside the region where the source is located,
$\rh$ becomes the usual Schwarzschild radius associated with the total 
Arnowitt-Deser-Misner (ADM)~\cite{adm} mass $M=m(\infty)$, 
\be
\Rh
=
2\,\lp\,\frac{M}{\mpl}
\ ,
\label{RhM}
\ee
and the above argument gives a mathematical foundation to
Thorne's {\em hoop conjecture\/}~\cite{Thorne:1972ji}.
\par
This description clearly becomes questionable for sources of the Planck size or
lighter, for which quantum effects may not be neglected.
The Heisenberg principle introduces an uncertainty in the spatial localisation
of the order of the Compton-de~Broglie length, $\lambda_M \simeq \lp\,{\mpl}/{M}$,
and we could argue that $\Rh$ only makes sense if
$\Rh\gtrsim \lambda_M$, that is $M \gtrsim \mpl$.
The HQM was precisely proposed in order to describe cases in which one expects quantum
uncertainties are not negligible.
For this purpose, we assume the existence of two observables, the quantum Hamiltonian
corresponding to the total energy $M$ of the system~\footnote{See also Ref.~\cite{baryons}
for further clarifications why $H$ is to be taken as the (super-)Hamiltonian of the ADM formalism.
We will return to this important point in Section~\ref{BECsection}.
\label{f5}},
\be
\hat H
=
\sum_\alpha
E_\alpha\ket{E_\alpha}\bra{E_\alpha}
\ ,
\label{HM}
\ee
where the sum is over the Hamiltonian eigenmodes, and the gravitational radius
with eigenstates
\be
\hat R_{\rm H}\,\ket{{\Rh}_\beta}
=
{\Rh}_\beta\,\ket{{\Rh}_\beta}
\ .
\ee
General states for our system can correspondingly be described by linear combinations of 
the form
\be
\ket{\Psi}
=
\sum_{\alpha,\beta}
C(E_\alpha,{\Rh}_\beta)\,
\ket{E_\alpha}
\ket{{\Rh}_\beta} 
\ ,
\label{Erh}
\ee
but only those for which the relation~\eqref{RhM} between the Hamiltonian and gravitational
radius holds are viewed as physical.
In particular, we impose~\eqref{RhM} after quantisation, as the weak Gupta-Bleuler 
constraint
\be
0
=
\left(
\hat H
-
\frac{\mpl}{2\,\lp}\,\hat R_{\rm H}
\right)
\ket{\Psi}
=
\sum_{\alpha,\beta}
\left(
E_\alpha-\frac{\mpl}{2\,\lp}\,{\Rh}_\beta
\right)
C(E_\alpha,{\Rh}_\beta)\,
\ket{E_\alpha}
\ket{{\Rh}_\beta} 
\ .
\label{Hcond}
\ee
The solution is clearly given by
\be
C(E_\alpha,{\Rh}_\beta)
=
C(E_\alpha,{2\,\lp}\,E_\alpha/{\mpl})\,\delta_{\alpha\beta}
\ ,
\label{solG}
\ee
which means that Hamiltonian eigenmodes and gravitational radius eigenmodes can only appear
suitably paired in a physical state.
The interpretation of this result is simply that the gravitational radius is not an independent degree
of freedom in our treatment, precisely because of the constraint~\eqref{RhM}~\footnote{For a comparison
with different approaches to horizon quantisation, see section~2.4 in Ref.~\cite{Casadio:2015qaq}.}.
\par
By tracing out the gravitational radius part, we recover the spectral decomposition
of the source wave-function,
\be
\ket{\psis}
&\!\!=\!\!&
\sum_\gamma\bra{{\Rh}_\gamma}\,
\sum_{\alpha,\beta}
\ket{{\Rh}_\beta}\, 
C(E_\alpha,{2\,\lp}\,E_\alpha/{\mpl})\,\delta_{\alpha\beta}
\ket{E_\alpha}
\nonumber
\\
&\!\!=\!\!&
\sum_\alpha
C\left(E_\alpha,{2\,\lp}\,E_\alpha/{\mpl}\right)
\ket{E_\alpha}
\nonumber
\\
&\!\!\equiv\!\!&
\sum_\alpha
C_{\rm S}(E_\alpha)\,\ket{E_\alpha}
\ ,
\ee
in which we used the (generalised) orthonormality of the gravitational radius eigenmodes~\cite{hqft}.
Note that the relation~\eqref{solG} now ensures that the result of this operation of integrating out
the gravitational radius is still a pure quantum state.
\par
Conversely, by integrating out the energy eigenstates, we will obtain the
Horizon Wave-Function (HWF)~\cite{fuzzyh,hqft}
\be
\ket{\psih}
&\!\!=\!\!&
\sum_\gamma\bra{E_\gamma}\,
\sum_{\alpha,\beta}
\ket{E_\alpha}\, 
C(E_\alpha,{2\,\lp}\,E_\alpha/{\mpl})\,\delta_{\alpha\beta}
\ket{{\Rh}_\beta}
\nonumber
\\
&\!\!=\!\!&
\sum_\alpha
C_{\rm S}({\mpl\,{\Rh}_\alpha}/{2\,\lp})
\ket{{\Rh}_\alpha}
\ ,
\ee
or
\be
\psih({\Rh}_\alpha)
=
\pro{{\Rh}_\alpha}{\psih}
=
C_{\rm S}({\mpl\,{\Rh}_\alpha}/{2\,\lp})
\ ,
\label{psihd}
\ee
where ${\mpl\,{\Rh}_\alpha}/{2\,\lp}=E({\Rh}_\alpha)$ is fixed by the constraint~\eqref{RhM}.
If the index $\alpha$ is continuous (again, see Ref.~\cite{hqft} for some important remarks),
the probability density that we detect a
gravitational radius of size $\Rh$ associated with the quantum state
$\ket{\psis}$ is given by $\mathcal{P}_{\rm H}(\Rh) = 4\,\pi\,\Rh^2\,|\psih(\Rh)|^2$,
and we can define the conditional probability density that the source lies
inside its own gravitational radius $\Rh$ as
\be
\mathcal{P}_<(r<\Rh)
=
P_{\rm S}(r<\Rh)\,\mathcal{P}_{\rm H}(\Rh)
\ ,
\label{PrlessH}
\ee
where $P_{\rm S}(r<\Rh)= 4\,\pi \int_0^{\Rh} |\psis(r)|^2\,r^2\,\d r$~\footnote{One
can also view $\mathcal{P}_<(r<\Rh)$ as the probability density that the sphere $r=\Rh$
is a horizon.}.
Finally, the probability that the system in the state $\ket{\psis}$ is a black hole
will be obtained by integrating~\eqref{PrlessH} over all possible values of $\Rh$,
namely
\be
P_{\rm BH}
=
\int_0^\infty
\mathcal{P}_<(r<\Rh)\,\d \Rh
\ .
\label{pbhgen}
\ee
Note that now the gravitational radius is necessarily ``fuzzy'' and characterised by
an uncertainty $\Delta\Rh = \sqrt{\expec{\Rh^2}-\expec{\Rh}^2}$.
\par
This quantum description for the total ADM mass $M$ and global gravitational radius $\Rh$
will be next extended to rotating sources by appealing to the asymptotic charges of
axially symmetric space-times.
We would like to recall that in Ref.~\cite{hqft} a local construction was also introduced based
on the quasi-local mass~\eqref{M}, which allows one to describe quantum mechanically any
trapping surfaces.
However, that local analysis cannot be extended to rotating sources without a better
understanding of the relation between quasi-local charges and the corresponding casual
structure~\cite{qlm-review}.
\subsection{Rotating sources: Kerr horizons}
\label{rotS}
Our aim is now to extend the HQM to rotating sources, for which there is no general consensus
about the proper quasi-local mass function to employ, and how to determine the causal structure 
from it.
For this reason, we shall explicitly consider relations that hold in space-times of the Kerr family,
generated by stationary axisymmetric sources which are both localised in space and subject
to a motion of pure rotation in the chosen reference frame.
\par
We assume the existence of a complete set of commuting operators
$\{\widehat H, \, \widehat{J} ^{2}, \, \widehat{J} _{z} \}$ acting on a Hilbert space
$\mathcal H$ connected with the quantum nature of the source. 
We also consider only the discrete part of the energy spectrum~\cite{hqft}, and
denote with $\alpha=\{a, \, j, \, m\}$ the set of quantum numbers parametrising
the spectral decomposition of the source, that is
\be
\label{eq-source-kerr}
\ket{\psi _{S}} = \sum _{a, j, m} C_{S} (E_{a \, j}, \lambda _{j} , \xi _{m}) \, \ket{a \, j \, m}
\ ,
\ee
where the sum formally represents the spectral decomposition in terms of the common
eigenmodes of the operators $\{\hat H, \, \hat{J} ^{2}, \, \hat{J} _{z} \}$.
In particular, we have that~\footnote{For later convenience, we rescale the standard
angular momentum operators $\hat\j^2$ and $\hat\j_z$ by factors of $\gn$
so as to have all operators proportional to $\mpl$ to a suitable power.}
\be
\hat H
&\!\!=\!\!&
\sum _{a, j , m} E_{a \, j} \, \ket{a \, j \, m} \bra{a \, j \, m}
\ ,
\\
\hat{J} ^{2} 
&\!\!\equiv\!\!&
\frac{\mpl^2}{\lp^2}\,\hat\j^2
=
\mpl^{4} \, \sum _{a, j , m} j \,(j+1) \, \ket{a \, j \, m} \bra{a \, j \, m} \equiv \sum _{a, j , m} \lambda _{j} \, \ket{a \, j \, m} \bra{a \, j \, m}
\ ,
\\
\hat{J} _{z} 
&\!\!\equiv\!\!&
\frac{\mpl}{\lp}\,\hat\j_z
=
\mpl^2 \,\sum _{a, j , m}  m \, \ket{a \, j \, m} \bra{a \, j \, m} \equiv \sum _{a, j , m} \xi _{m} \, \ket{a \, j \, m} \bra{a \, j \, m}
\ .
\ee
From the previous discussion, one can also easily infer that
$j \in \mathbb{N} _{0} /2$, $m \in \mathbb{Z} / 2$, with $|m| \leq j$,
and $a \in \mathcal{I}$, where $\mathcal{I}$ is a \textit{discrete}
set of labels that can be either finite of infinite.
\par 
Let us first note that Eq.~\eqref{eq-source-kerr} stems from the idea that the
space-time should reflect the symmetries of the source.
Therefore, our first assumption is that the source should obviously have an angular momentum
in order to describe a rotating black hole.
Now, for a stationary asymptotically flat space-time, we can still define the ADM mass
$M$ and, following Ref.~\cite{hqft} as outlined in the previous subsection, we can replace this
classical quantity with the expectation value of our Hamiltonian~\footnote{See footnote~\ref{f5}.},
\be
M
\to
\bra{\psis} \hat H \ket{\psis}
&\!\!=\!\!&
\sum _{a, j, m}
\sum _{b, k, n}
C^{\ast} _{S} (E_{a \, j}, \lambda _{j} , \xi _{m}) \,
C_{S} (E_{b \, k}, \lambda _{k} , \xi _{n}) \, 
\bra{a \, j \, m} \widehat H \ket{b \, k \, n}
\notag
\\
&\!\!=\!\!& 
\sum _{a, j, m}
|C _{S} (E_{a \, j}, \lambda _{j} , \xi _{m})|^{2} \, E_{a \, j}
\ .
\ee
In General Relativity, we can also define a conserved classical charge 
arising from the axial symmetry by means of the Komar integral.
This will be the total angular momentum $J$ of the Kerr spacetime.
However, in our description of the quantum source, we have two distinct notions
of angular momentum, i.e.~the total angular momentum
\be
\bra{\psis} \hat{J} ^{2} \ket{\psis}
=
\sum _{a, j, m} |C _{S} (E_{a \, j}, \lambda _{j} , \xi _{m})|^{2} \, \lambda_{j}
\ ,
\ee
and the component of the angular momentum along the axis of symmetry
\be
\bra{\psis} \hat{J} _{z} \ket{\psis}
=
\sum _{a, j, m} |C _{S} (E_{a \, j}, \lambda _{j} , \xi _{m})|^{2} \, \xi_{m}
\ .
\ee
Since, at least classically, we can always rotate our reference frame so that 
the axis of symmetry is along the $z$ axis, it is reasonable to consider
$\hat{J}^2$ as the quantum extension of the classical angular momentum
for a Kerr black hole, 
\be
J^{2} \to \bra{\psis} \hat{J} ^{2} \ket{\psis}
=
\sum _{a, j, m} |C _{S} (E_{a \, j}, \lambda _{j} , \xi _{m})|^{2} \, \lambda_{j}
\ .
\ee 
In the following, we will further assume that $\expec{\hat{J}_{z}}$
is maximum in our quantum states,
so that the proper (semi-)classical limit is recovered,
that is
\be 
\bra{\psis}\left( \hat{J} ^{2} - \hat{J} _{z}^2\right) \ket{\psis}
\
\underset{\hbar\to 0}{\overset{j \to \infty}{\longrightarrow}}
\
0
\ ,
\label{JJz}
\ee
for $\hbar\,j=\lp\,\mpl\,j$ held constant.
\par
For the Kerr space-time we have two horizons given by
\be
\Rh^{(\pm)}
=
\frac{\lp}{\mpl}
\left(
M \pm \sqrt{M^{2} - \frac{J^{2}}{M^{2}}}
\right)
\ ,
\label{R+-}
\ee
provided $J^2<M^4$. 
Let us then introduce two operators $\hat R ^{(\pm)}$
and, for the sake of brevity, write their eigenstates as
\be
\hat R ^{(\pm)} _{\rm H}\,\ket{\beta} _{\pm}
=
{\Rh}^{(\pm)} _\beta\,\ket{\beta} _{\pm}
\ .
\ee
The generic state for our system can now be described by a triply entangled
state given by
\be
\ket{\Psi}
=
\sum_{a, j , m} \sum _{\alpha , \beta}
C(E_{a\, j}, \lambda _{j} , \xi _{m}, {\Rh} ^{(+)} _\alpha , {\Rh} ^{(-)} _\beta) \,
\ket{a \, j \, m}
\ket{\alpha}_{+}
\ket{\beta} _{-} 
\ ,
\label{Erh}
\ee
but Eq.~\eqref{R+-} tells us that in order to be able to define the analogue of the
condition~\eqref{Hcond} for the rotating case, we have to assume some mathematical
restrictions on the operator counterparts of $M$ and $J$.
First of all, the term $J^{2} / M^{2}$ tells us that we should assume $\hat H$ to be
an \textit{invertible self-adjoint} operator, so that
\be
J^{2} / M^{2}
\
\to
\ 
\hat{J} ^{2} \, (\hat{H} ^{-1}) ^{2}
=
(\hat{H} ^{-1}) ^{2} \,\hat{J} ^{2}
\ .
\ee
For this purpose, it is useful to recall a corollary of the spectral theorem:
\begin{corollary}
Let $\hat A$ be a self-adjoint positive semi-definite operator.
Then $\hat A$ has a positive semi-definite square root $\hat S$, that is, $\hat S$ is self-adjoint,
positive semi-definite, and
$$
\hat{S}^{2}=\hat A
\ . 
$$
If $\hat A$ is positive definite, then $\hat S$ is positive definite.
\end{corollary}
\noindent
It follows that the operator $\hat{H} ^{2} - \hat{J} ^{2} \, (\hat{H} ^{-1}) ^{2}$ should be, at least,
a positive semi-definite operator.
On defining the operators
\be
\hat{\mathcal{O}} ^{\pm}
\equiv
\hat H \pm \left( \hat{H} ^{2} - \hat{J} ^{2} \, \hat{H} ^{-2} \right) ^{1/2}
\, ,
\ee 
we obtain that the physical states of the system are those simultaneously satisfying
\be
\left(\hat R_{\rm H} ^{(+)} - \hat{\mathcal{O}} ^{+} \right)
\ket{\Psi} _{\rm phys}
=
0
\label{hR+}
\ee
and
\be
\left(\hat R_{\rm H} ^{(-)} - \hat{\mathcal{O}} ^{-}\right)
\ket{\Psi} _{\rm phys}
=
0
\ .
\label{hR-}
\ee 
These two conditions reduce to
\be
C(\{a \, j \, m\}, {\Rh} ^{(+)} _\alpha , {\Rh} ^{(-)} _\beta)
&\!\!=\!\!&
C(E_{a\, j}, \{j \, m\}, {\Rh} ^{(+)} _{a\, j} (E _{a\, j}) , {\Rh} ^{(-)} _\beta) \, \delta _{\alpha , \{ a, j \}}
\ ,
\\
C(\{a \, j \, m\}, {\Rh} ^{(+)} _\alpha , {\Rh} ^{(-)} _\beta)
&\!\!=\!\!&
C(E_{a\, j}, \{j \, m\}, {\Rh} ^{(+)} _\alpha , {\Rh} ^{(-)} _{a\, j} (E _{a\, j})) \, \delta _{\beta , \{ a, j \}}
\ ,
\ee
from which we obtain
\be
C(E_{a\, j}, \{j \, m\}, {\Rh} ^{(+)} _\alpha , {\Rh} ^{(-)} _\beta)
=
C(E_{a\, j}, \{j \, m\}, {\Rh} ^{(+)} _{a\, j} (E _{a\, j}) , {\Rh} ^{(-)} _{a\, j} (E _{a\, j})) \,
\delta _{\alpha , \{ a, j \}} \, \delta _{\beta , \{ a, j \}}
\ .
\ee	
By tracing out the geometric parts, we should recover the matter state, that is
\be
\ket{\psis}
=
\sum_{a, j , m} C(E_{a\, j}, \lambda_{j}, \xi_{m}, {\Rh} ^{(+)} _{a\, j} (E _{a\, j}) , {\Rh} ^{(-)} _{a\, j} (E _{a\, j})) \, \ket{a \, j \, m}
\ ,
\ee 
which implies
\be
C_{S} (E_{a\, j}, \lambda_{j}, \xi_{m})
=
C(E_{a\, j}, \lambda_{j}, \xi_{m}, {\Rh} ^{(+)} _{a\, j} (E _{a\, j}) , {\Rh} ^{(-)} _{a\, j} (E _{a\, j}))
\ .
\ee 
Now, by integrating away the matter state, together with one of the two geometric parts,
we can compute the wave function corresponding to each horizon,
\be
\psi_\pm(\Rh^{(\pm)})
=
C(E _{a\, j}(\Rh^{(\pm)}), \lambda_{j}(\Rh^{(\pm)}), \xi_{m}(\Rh^{(\pm)}))
\ .
\label{psi+-}
\ee
It is also important to stress that the hamiltonian constraints imply a strong relation between
the two horizons, indeed we have that $R_{H} ^{\pm} = R_{H} ^{\pm} (R_{H} ^{\mp})$.
\section{Corpuscular Harmonic Black Holes}
\setcounter{equation}{0}
\label{BECsection}
In the corpuscular model proposed by Dvali and Gomez~\cite{dvali}, black holes are macroscopic
quantum objects made of gravitons with a very large occupation number $N$ in the ground-state,
effectively forming Bose Einstein Condensates. 
As also derived in Ref.~\cite{baryons} from a post-Newtonian analysis of the coherent state
of gravitons generated by a matter source, the virtual gravitons forming the black hole of radius
$\Rh$ are ``marginally bound'' by their Newtonian potential energy $U$, that is
\be
\mu + U_{\rm N}
\simeq
0
\ ,
\label{energy0}
\ee
where $\mu$ is the graviton effective mass related to their quantum mechanical
size via the Compton/de~Broglie wavelength $\lambda_\mu\simeq\lp\,{\mpl}/{\mu}$,
and $\lambda_\mu\simeq\Rh$.
\par
A first rough approximation for the potential energy $U_{\rm N}$ is obtained by considering a 
square well for $r<\lambda_\mu$,
\be
U
\simeq
-N\,\alpha\,\frac{ \hbar}{\lambda_\mu}\,\Theta(\ell-r)
\ ,
\label{Udvali}
\ee
where $\Theta$ is the Heaviside step function and the coupling constant 
$\alpha={\lp^2}/{\lambda_\mu^2}={\mu^2}/{\mpl^2}$.
The energy balance~\eqref{energy0} then leads to $N\,\alpha = 1$ and,
with $\lambda_\mu\simeq\Rh$,
\be
\mu
\simeq
\mpl\,\frac{\lp}{\Rh}
\simeq
\frac{\mpl}{\sqrt{N}}
\ ,
\ee
so that
\be
M
\simeq
N\,\mu
\simeq
\sqrt{N}\,\mpl
\ .
\label{Max}
\ee
\par
A better approximation for the potential energy was employed in Ref.~\cite{QHBH}, which
takes the harmonic form
\be
V
&\!\!=\!\!&
\frac{1}{2}\,\mu\,\omega^2\,(r^2-d^2)\,\Theta(d-r)
\nonumber
\\
&\!\!\equiv\!\!&
V_0(r)\,\Theta(d-r)
\ ,
\label{Vho}
\ee
where the parameters $d$ and $\omega$ will have to be so chosen as to ensure the highest
energy mode available to gravitons is just marginally bound [see Eq.~\eqref{energy0}].
If we neglect the finite size of the well, the Schr\"odinger equation in spherical coordinates,
\be
\frac{\hbar^2}{2\,\mu\,r^2}
\left[
\frac{\partial}{\partial r} \left( r^2 \,\frac{\partial}{\partial r} \right)
+
\frac{1}{\sin \theta}\, \frac{\partial}{\partial \theta}
\left(  \sin \theta\, \frac{\partial}{\partial \theta} \right)
+ \frac{1}{\sin^2 \theta} \,\frac{\partial^2}{\partial \phi^2}
\right]
\psi
=
(V_0-\en)\,
\psi
\ ,
\label{schrod}
\ee
yields the well-known eigenfunctions
\be
\psi_{njm}(r,\theta,\phi;\lambda_\mu)
=
\mathcal{N}\,r^l\, e^{-\frac{r^2}{ 2\,\lambda_\mu^2}}\, _1F_1(-n,l+3/2,r^2/\lambda_\mu^2)\, Y_{lm}(\theta,\phi)
\ ,
\label{psin}
\ee
where $\mathcal{N}$ is a normalization constant, $_1F_1$ the Kummer confluent
hypergeometric function of the first kind and $Y_{lm}(\theta,\phi)$
are the usual spherical harmonics.
The corresponding eigenvalues are given by
\be
\en_{nl}
=
\hbar\, \omega
\left(2\,n+l+ \frac{3}{2}\right)
+V_0(0)
\ ,
\label{Enl}
\ee
where $n$ is the radial quantum number.
It is important to remark that the quantum numbers $l$ and $m$ here must not be confused
with the total angular momentum numbers $j$ and $m$ of Section~\ref{rotS},
as the latter are the sum of the former.
At the same time, the ``energy'' eigenvalues $\en_{nl}$ must not be confused with the ADM
energy $E_{aj}$ of that Section, here equal to $N\,\mu$ by construction.
\par
If we denote with $n_0$ and $l_0$ the quantum numbers of the
highest ``energy'' state, 
and include the graviton effective mass $\mu$ in the constant $V_0(0)$,
the condition~\eqref{energy0} becomes $\en_{n_0 l_0}\simeq 0$, or
\be
V_0(0)
\simeq
-\hbar\, \omega
\left(2\,n_0+ l_0 + \frac{3}{2}\right)
\ ,
\ee
which yields
\be
\omega\,d^2
\simeq
2\,\frac{\hbar}{\mu}
\left(2\,n_0+ l_0 + \frac{3}{2}\right)
\ .
\ee
We now further assume that $d\simeq\lambda_\mu\simeq\Rh^{(+)}$ and use the Compton
relation for $\mu$, so that the above relation fully determines 
\be
\omega
\simeq
\frac{2}{\lambda_\mu}
\left(2\,n_0+ l_0 + \frac{3}{2}\right)
\ .
\ee
The potential can be finally written as
\be
V_0
=
2\,\mu
\left(2\,n_0+ l_0 + \frac{3}{2}\right)^2
\frac{r^2-\lambda_\mu^2}{\lambda_\mu^2}
\ee
and the eigenvalues as
\be
\en_{nl}
&\!\!\simeq\!\!&
-\hbar\,\omega
\left[2\,(n_0-n)+(l_0-l)\right]
\nonumber
\\
&\!\!\simeq\!\!&
-2\,\mu
\left(2\,n_0+ l_0 + \frac{3}{2}\right)
\left[2\,(n_0-n)+( l_0-l)\right]
\nonumber
\\
&\!\!\equiv\!\!&
-2\,\mu_0
\left[2\,(n_0-n)+( l_0-l)\right]
\ ,
\ee
which of course holds only for $n\le n_0$ and $l\le l_0$.
Let us remark that the fact the above ``energy'' is negative for the allowed
values of $n$ and $j$ is indeed in agreement with the post-Newtonian analysis of the
``maximal packing condition'' for the virtual gravitons in the black hole~\cite{baryons}~\footnote{It
becomes positive if we consider the effective mass $\mu<0$ for virtual gravitons.}.
\par
In the following, we shall consider a few specific states in order to show the kind of
results one can obtain from the general HQM formalism of Section~\ref{rotS} applied to 
harmonic models of spinning black holes.
\subsection{Rotating Black Holes}
We shall now consider some specific configurations of harmonic black holes with
angular momentum and apply the extended HQM described in the previous section.
We first remark that the quantum state of $N$ identical gravitons will be a $N$-particle
state, i.e.~a vector of the $N$-particle Fock space $\mathcal{F} = \mathcal{H} ^{\otimes N}$,
where $\mathcal{H}$ is a suitable 1-particle Hilbert space. 
However, both the Hamiltonian of the system $\hat{H}$ and the gravitational radius
$\hat R_{\rm H}$ are global observables and act as $N$-body operators on $\mathcal{F}$.
\subsubsection{Single eigenstates}
The simplest configuration corresponds to all toy gravitons in the same mode,
and the quantum state of the system is therefore given by
\be 
\ket{\Psi} \equiv \ket{M \, J} = \bigotimes _{\alpha = 1} ^N \ket{g} _\alpha
\ ,
\ee
where $\ket{g}$ represents the wave-function of a single component.
In particular, this $\ket{\Psi}$ is a Hamiltonian eigenstate, for which the total ADM energy
is simply given by
\be
\bra{\Psi}\hat H\ket{\Psi}
\equiv
\expec{\hat H}
=
N\,\mu
=
M
\ ,
\ee
and each graviton is taken in one of the modes~\eqref{psin}.
For the sake of simplicity, we shall set $n=n_0=0$, $l=l_0=2$ and $m=\pm 2$, that is
\be
\expec{r, \theta, \phi \, | \, g} =
\psi_{02\pm2} (r, \theta, \phi;\lambda_\mu)
=
\mathcal{N}\,r^2\,\exp\left(- \frac{r^2}{2\,\lambda_\mu^2} \right) \,Y_{2\pm2}(\theta,\phi)
\ ,
\label{psi022}
\ee
where the normalisation constant $ \mathcal{N} = {4}/({\sqrt{15} \, \pi ^{1/4} \, \lambda _\mu ^{7/2}})$.
The total angular momentum is thus given by 
\be
\bra{\Psi}\hat J^2\ket{\Psi}
\equiv
\expec{\hat{J} ^2}
&\!\!=\!\!&
4\,(N_+-N_-)(N_+-N_-+1/2)\,\mpl^2
\nonumber
\\
&\!\!\equiv\!\!&
4\,L^2\,N^2\,\mpl^2
\ ,
\ee
where $N_+\ge N_-=N-N_+$ is the number of spin up constituents (with $m=+2$).
We also introduced the constant
\be
L^2
&\!\!=\!\!&
\left(\frac{N_+}{N}-\frac{N_-}{N}\right)\left(\frac{N_+}{N}-\frac{N_-}{N}+\frac{1}{2\,N}\right)
\nonumber
\\
&\!\!\equiv\!\!&
\left(2\,n_+-1\right)\left(2\,n_+-1+\frac{1}{2\,N}\right)
\simeq
\left(2\,n_+-1\right)^2
\ ,
\ee
where the approximate expression holds for $N\gg 1$.
Note that $L^2=0$ for $n_+\equiv N_+/N=1/2$ (the non-rotating case with $N_+=N_-$) and grows
to a maximum $L^2\simeq 1+\mathcal{O}(1/N)$ for the maximally rotating case $n_+=1$ (or $N_+=N$).
\par
Since we are considering an eigenstate of both the Hamiltonian $\hat H$ and total
angular momentum $\hat J^2$, the wave-functions~\eqref{psi+-} for the two horizons 
will reduce to single eigenstates of the respective gravitational radii as well.
In particular, replacing the above values into~\eqref{hR+} and~\eqref{hR-} yields
\be
\expec{\hat R_{\rm H}^{(\pm)}}
=
N\,\lp\,\frac{\mu}{\mpl}
\left(
1
\pm
\sqrt{1
-
4\,\frac{L^2\,\mpl^4}{N^2\,\mu^4}}
\right)
\ .
\label{expRh}
\ee
The classical condition for the existence of these horizons is that the square root be real,
which implies
\be
\mu^2
\ge
2\,\mpl^2\,\frac{L}{N}
\ .
\ee
The above bound vanishes for $N_+=N_-=N/2$, as expected for a spherical black hole, 
and is maximum for $N_+=N$, in which case it yields
\be
\mu^2
\gtrsim 
2\,\frac{\mpl^2}{N}
\ ,
\ee
again for $N\gg 1$.
\begin{figure}[t]
\centering
\includegraphics[width=10cm]{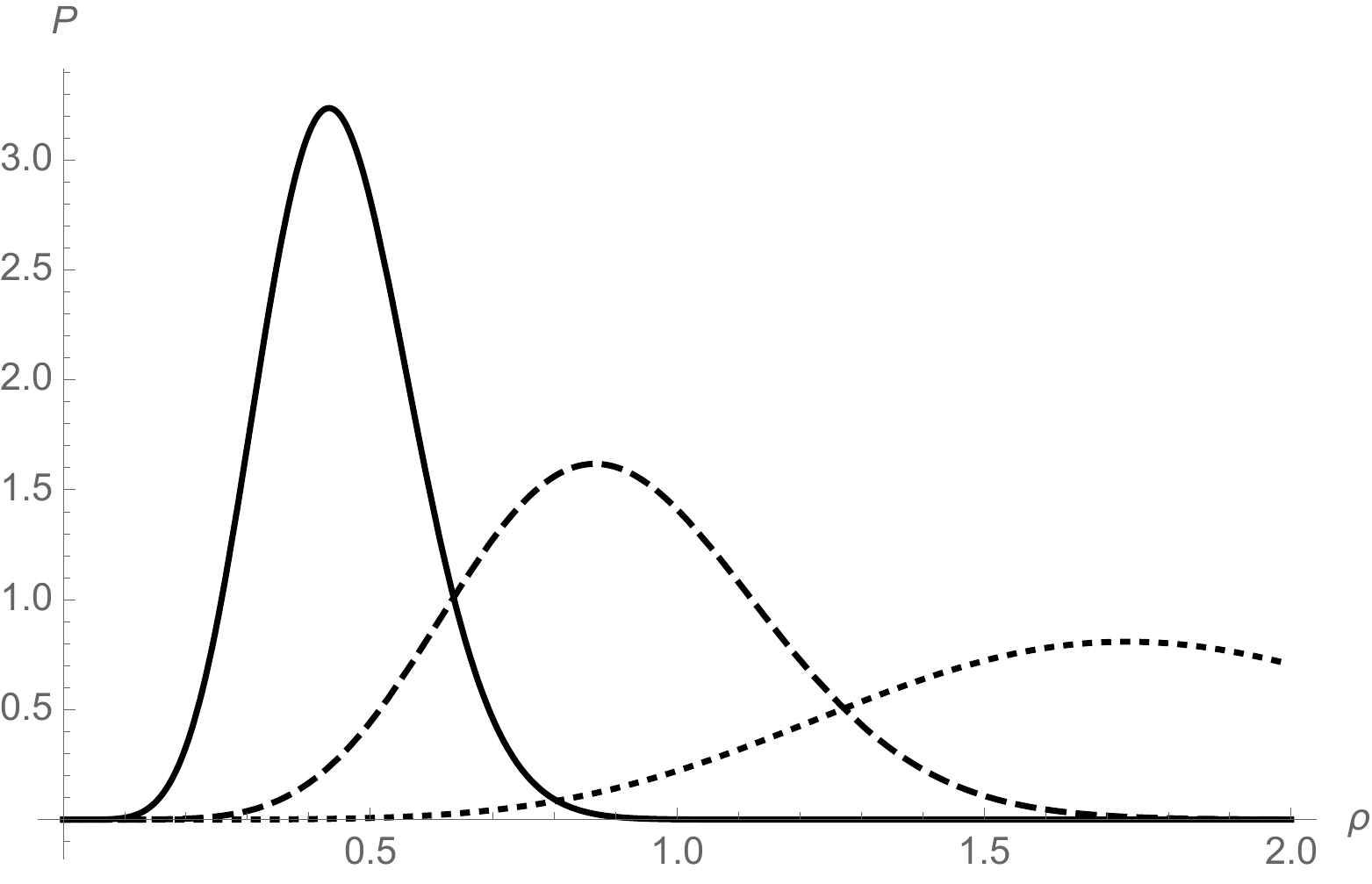}
\caption{Plots of ${\mathcal P}_{02}(\rho;\lambda_\mu)$ as a function of
$\rho=r/\expec{\hat R_{\rm H}^{(+)}}$ for
$\lambda_\mu= \expec{\hat R_{\rm H}^{(+)}}/4$ (solid line),
$\lambda_\mu= \expec{\hat R_{\rm H}^{(+)}}/2$ (dashed line)
and $\lambda_\mu= \expec{\hat R_{\rm H}^{(+)}}$ (dotted line).}
\label{Pr}
\end{figure}
\par
Since we are modelling black holes, it is particularly interesting to study in details the consequences
of assuming that all the constituents of our system lie inside the outer horizon.
In other words, we next require that the Compton length of gravitons, $\lambda_\mu=\lp\,\mpl/\mu$,
is such that the modes~\eqref{psi022} are mostly found inside the outer horizon radius
$\expec{\hat R_{\rm H}^{(+)}}$.	
In order to impose this condition, we compute the single-particle probability density 
\be
\mathcal{P}_{02}(r;\lambda_\mu)
=
\int_{-1}^{+1} \d \cos\theta
\int_0^{2\,\pi} \d\phi
\left| \psi_{02+2}(r,\theta,\phi;\lambda_\mu) \right| ^2
=
\mathcal{N}^2 \, r^6 \, \exp \left( - \frac{r ^2}{\lambda _\mu ^2} \right)
\ , 
\label{single-density}
\ee
where we used $|\psi_{02+2}|^2=|\psi_{02-2}|^2$.
From Fig.~\ref{Pr}, we then see that this probability is peaked well inside
$\expec{\hat R_{\rm H}^{(+)}}$ for $\lambda_\mu=\expec{\hat R_{\rm H}^{(+)}}/4$,
whereas $\lambda_\mu=\expec{\hat R_{\rm H}^{(+)}}/2$ is already borderline
and $\lambda_\mu=\expec{\hat R_{\rm H}^{(+)}}$ is clearly unacceptable.
\par
We find it in general convenient to introduce the variable
\be
\gamma(n_+,N)
\equiv
\frac{\expec{\hat R_{\rm H}^{(+)}}}{2\,\lambda_\mu}
\ ,
\label{gamma}
\ee
which should be at least $1$ according to the above estimate, 
so that Eq.~\eqref{expRh} reads
\be
\frac{2\,\gamma\,\mpl^2}{N\,\mu^2}
\simeq
1
+
\sqrt{1
-
\frac{4\,L^2\,\mpl^4}{N^2\,\mu^4}}
\ ,
\ee
which we can solve for $x=2\,\gamma\,\mpl^2/(N\,\mu^2)$, that is
\be
(x-1)^2
\simeq
1-
{\ell^2}\,x^2
\ ,
\label{eqx}
\ee
with the condition $\ell\,x\equiv (L/\gamma)\,x\le 1$ to ensure the existence of the square root.
The only positive solution is given by
\be
x
\simeq
\frac{2}{1+\ell^2}
\ ,
\ee
for which the existence condition reads $(\ell-1)^2\ge 0$ and is identically satisfied.
The effective mass is then given by
\be
\mu^2
=
\frac{2\,\gamma\,\mpl^2}{N\,x}
\simeq
\frac{\gamma^2+L^2}{\gamma\,N}
\,\mpl^2
\ .
\label{mu2}
\ee
As a function of $N/2\le N_+\le N$, the above squared mass interpolates almost
linearly between $\mu_0^2=\gamma\,\mpl^2/N$ for $N_+=N_-=N/2$ (so that $L^2=0$)
and $\bar \mu^2\simeq (1+\gamma^2)\,{\mpl^2}/(\gamma\,N)$ for the maximally rotating case case
$N_+=N\gg 1$ (for which $L^2\simeq 1$).
The Compton length reads
\be
\lambda_{\mu}
=
\lp\,\frac{\mpl}{\mu}
\simeq
\sqrt{\frac{\gamma\,N}{\gamma^2+L^2}}\,\lp
\ ,
\ee
the ADM mass is
\be
M
=
N\,\mu
\simeq
\sqrt{\frac{\gamma^2+L^2}{\gamma}\,N}\,\mpl
\ .
\label{pureH}
\ee
and the angular momentum
\be
\expec{\hat J^2}
\simeq
4\,N^2\,L^2\,\mpl^4
\simeq
\frac{4\,\gamma^2\,L^2\,M^4}{(\gamma^2+L^2)^2}
<M^4
\ ,
\ee
for all values of $L\ge 0$.
This seems to suggest that $N$ constituents of effective mass $\mu\sim\mpl/\sqrt{N}$ 
cannot exceed the classical bound for black holes, or that naked singularities cannot be associated
with such multi-particle states.
However, a naked singularity has no horizon and we lose the condition~\eqref{energy0} from which 
the effective mass $\mu$ is determined.
If naked singularities can still be realised in the quantum realm, they must be described in a
qualitatively different way from the present one~\footnote{See Refs.~\cite{Casadio:2015rwa}
for spherically symmetric charged sources.}. 
\par
\begin{figure}[t]
\centering
\includegraphics[width=10cm]{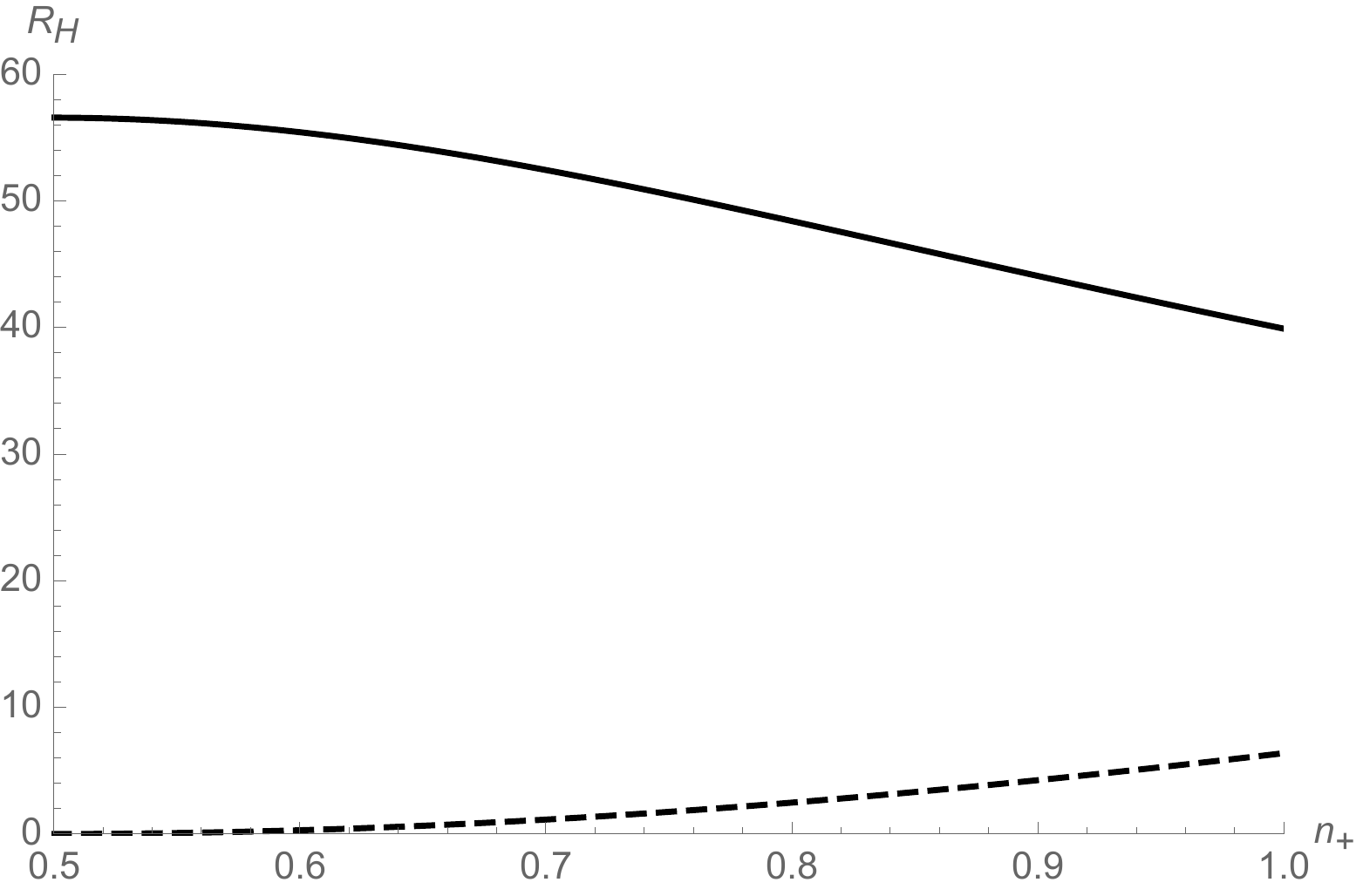}
\caption{Horizon radius $\expec{\hat R_{\rm H}^{(+)}}$ (solid line) and $\expec{\hat R_{\rm H}^{(-)}}$
(dashed line) in Planck length units for $N=100$ and $\gamma=2$.}
\label{RHpm2}
\end{figure}

Let us now replace the effective mass~\eqref{mu2} into Eq.~\eqref{expRh},
\be
\expec{\hat R_{\rm H}^{(\pm)}}
&\!\!\simeq\!\!&
\lp\,\sqrt{\frac{\gamma^2+L^2}{\gamma}\,N}
\left(
1
\pm
\sqrt{1
-
\frac{4\,\gamma^2\,L^2}{(\gamma^2+L^2)^2}}
\right)
\nonumber
\\
&\!\!\simeq\!\!&
\lp\,\sqrt{\frac{N/\gamma}{\gamma^2+L^2}}
\left(
\gamma^2+L^2
\pm
\left|\gamma^2-L^2\right|
\right)
\ .
\label{expRhL}
\ee
One has $L^2=\gamma^2$ for 
\be
n_+
=
n_{\rm c}
\equiv
\frac{2\,N-1+\sqrt{1+4\,\gamma^2\,N^2}}{4\,N}
\simeq
\frac{1+\gamma}{2}
-\frac{1}{4\,N}
\ .
\ee
Since $1/2\le n_+\le 1$, the critical value $n_{\rm c}$ becomes relevant only
for $\gamma\simeq N\simeq 1$.
For $N\gg 1$ and $\gamma\gtrsim 1$, the horizon radii are thus given by 
\be
\expec{\hat R_{\rm H}^{(-)}}
\simeq
\frac{L^2}{\gamma^2}\,\expec{\hat R_{\rm H}^{(+)}}
\simeq
2\,L^2\,\lp\,\sqrt{\frac{N/\gamma}{\gamma^2+L^2}}
\ ,
\label{1rh-}
\ee
and $2\,\gamma\,\lambda_\mu\simeq \expec{\hat R_{\rm H}^{(+)}}$, as we required.
The above horizon structure for $1/2\le n_+\le 1$ is displayed for $\gamma=2$ 
and $N=100$ in Fig.~\ref{RHpm2}, where we also remind that
$\lambda_\mu=\expec{\hat R_{\rm H}^{(+)}}/4$.
\par
\begin{figure}[t]
\centering
\includegraphics[width=10cm]{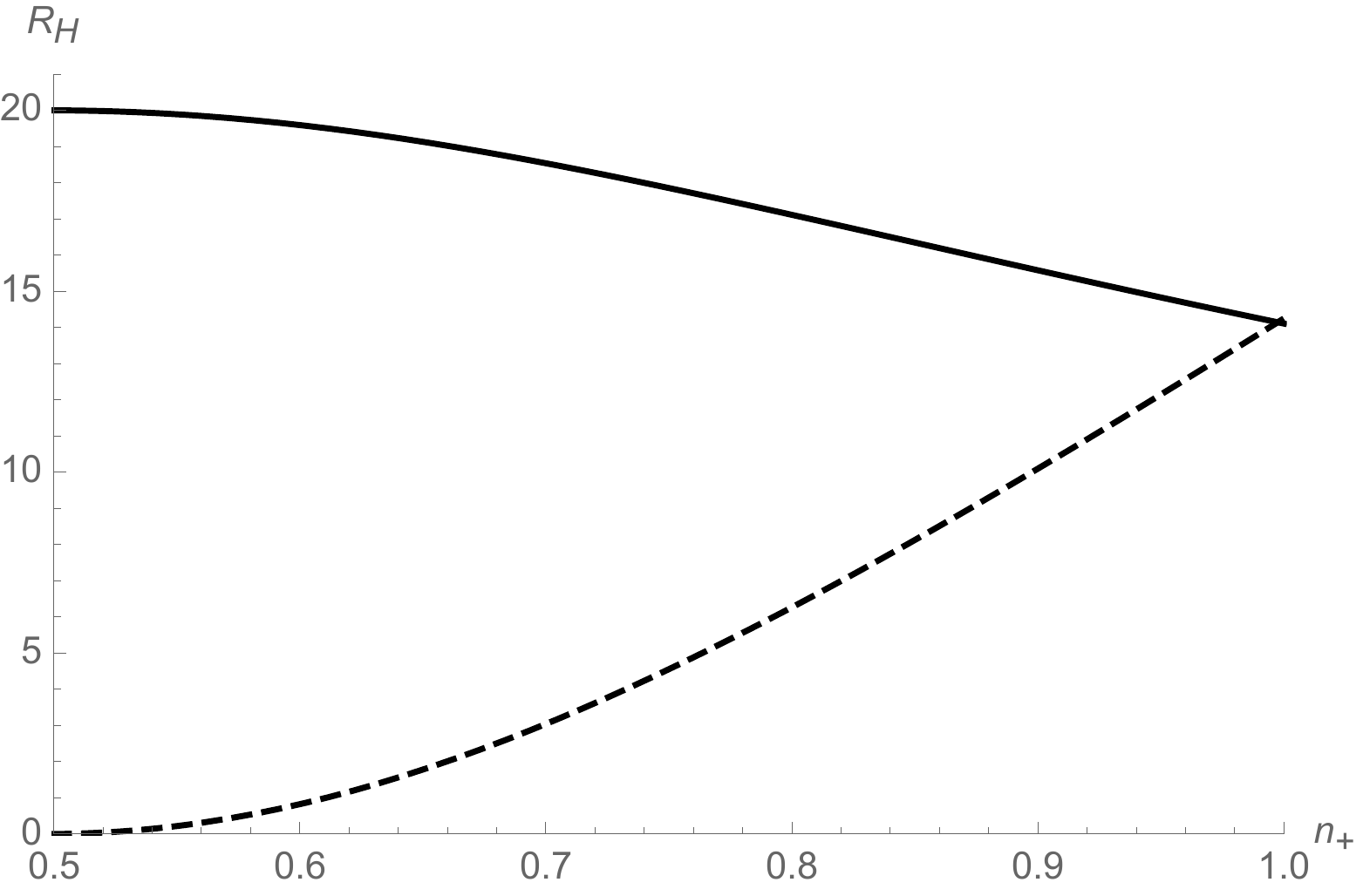}
\caption{Horizon radius $\expec{\hat R_{\rm H}^{(+)}}$ (solid line) and $\expec{\hat R_{\rm H}^{(-)}}$
(dashed line) in Planck length units for $N=100$ and $\gamma=1$.}
\label{RHpm}
\end{figure}

It is particularly interesting to note that the extremal Kerr geometry can only be realised 
in our model if $\gamma$ is sufficiently small.
In fact, $\expec{\hat R_{\rm H}^{(-)}}\simeq\expec{\hat R_{\rm H}^{(+)}}$ requires
\be
\gamma^2
\simeq
L^2
\ .
\ee
For $\gamma=1$ and $N=100$, the horizon structure is displayed in Fig.~\ref{RHpm}, 
where we see that the two horizons meet at $L^2\simeq 1$, that is the configuration
with $n_+\simeq 1$ in which (almost) all constituents are aligned.
Note also that, technically, for $N\simeq 1$ and $\gamma$ small, there would be a finite range
$n_{\rm c}<n_+\le 1$ in which the expressions of the two horizon radii switch.
However, this result is clearly more dubious as one would be dealing with a truly quantum
black hole made of a few constituents just loosely confined.
Such configurations could play a role in the formation of black holes, or in the final stages
of their evaporation, but we shall not consider this possibility any further here.
\par
Finally, let us apply the HQM and compute the probability~\eqref{pbhgen} that the system
discussed above is indeed a black hole.
We first note that, since we are considering eigenstates of the gravitational radii,
the wave-function~\eqref{psi+-} for the outer horizon will just contribute a Dirac delta peaked
on the outer expectation value~\eqref{expRh} to the general expression~\eqref{PrlessH},
that is
\be
{\mathcal{P}}_{(+)}(\Rh)
=
\delta(\Rh -\expec{\hat{R} ^{(+)} _{\rm H}})
\ .
\label{ph1m}
\ee
This implies
\be 
P_{\rm BH}(n_+, N)
=
P_<^{(+)} (r_1 < \expec{\hat{R} ^{(+)} _{\rm H}},\ldots,r_N < \expec{\hat{R} ^{(+)} _{\rm H}})
\ .
\ee
Moreover, since
\be
\expec{\textbf{r} _1, \ldots , \textbf{r} _N \, | \, \Psi}
=
\prod _{\alpha = 1} ^N \expec{\textbf{r}_\alpha \, | \, g} _\alpha
=
\prod _{\alpha = 1} ^N \psi_{02\pm2} (r _\alpha, \theta _\alpha, \phi _\alpha;\lambda_\mu)
\ ,
\ee
where $\textbf{r} \equiv (r, \, \theta, \, \phi)$, the joint probability density in position space is simply
given by
\be
\mathcal{P}(r_1,\ldots, r_N; \lambda _\mu ) 
=
\prod _{\alpha = 1} ^N \mathcal{P}_{02}(r_\alpha; \lambda _\mu)
=
\mathcal{N} ^{2N}
\, r_1 ^6 \cdots r_N ^6 \, \exp \left( - \frac{r_1 ^2 + \cdots + r_N ^2}{\lambda _\mu ^2} \right)
\ ,
\ee
where we used Eq.~\eqref{single-density}.
It immediately follows that 
\be
P_{\rm BH} (n_+ , N) 
=
\prod _{\alpha = 1}^N
P_< ^{(+)} (r_\alpha < \expec{\hat{R} ^{(+)} _{\rm H}})
=
\left[P_< ^{(+)} (r < \expec{\hat{R} ^{(+)} _{\rm H}})\right]^N
\ ,
\label{1mpbh}
\ee
with
\be
P_< ^{(+)}(r< \expec{\hat{R} ^{(+)} _{\rm H}})
&\!\!=\!\!&
\int_0^{\expec{\hat{R} ^{(+)} _{\rm H}}}
\mathcal{P}_{02}(r;\lambda_\mu)\, \d r
\nonumber
\\
&\!\!=\!\!&
{\rm erf}\!\left(2\,\gamma\right)
-\frac{4}{15\,\sqrt{\pi}}\,\gamma
\left(
64\,\gamma^4
+40\,\gamma^2
+15
\right)
e^{-4\,\gamma^2}
\nonumber
\\
&\!\!\equiv\!\!&
P_{(+)}(\gamma)
\ ,
\label{1pbh}
\ee
where we recall $\gamma$ was defined in Eq.~\eqref{gamma}, and depends on $N$ and $n_+$.
\begin{figure}[t]
\centering
\includegraphics[width=10cm]{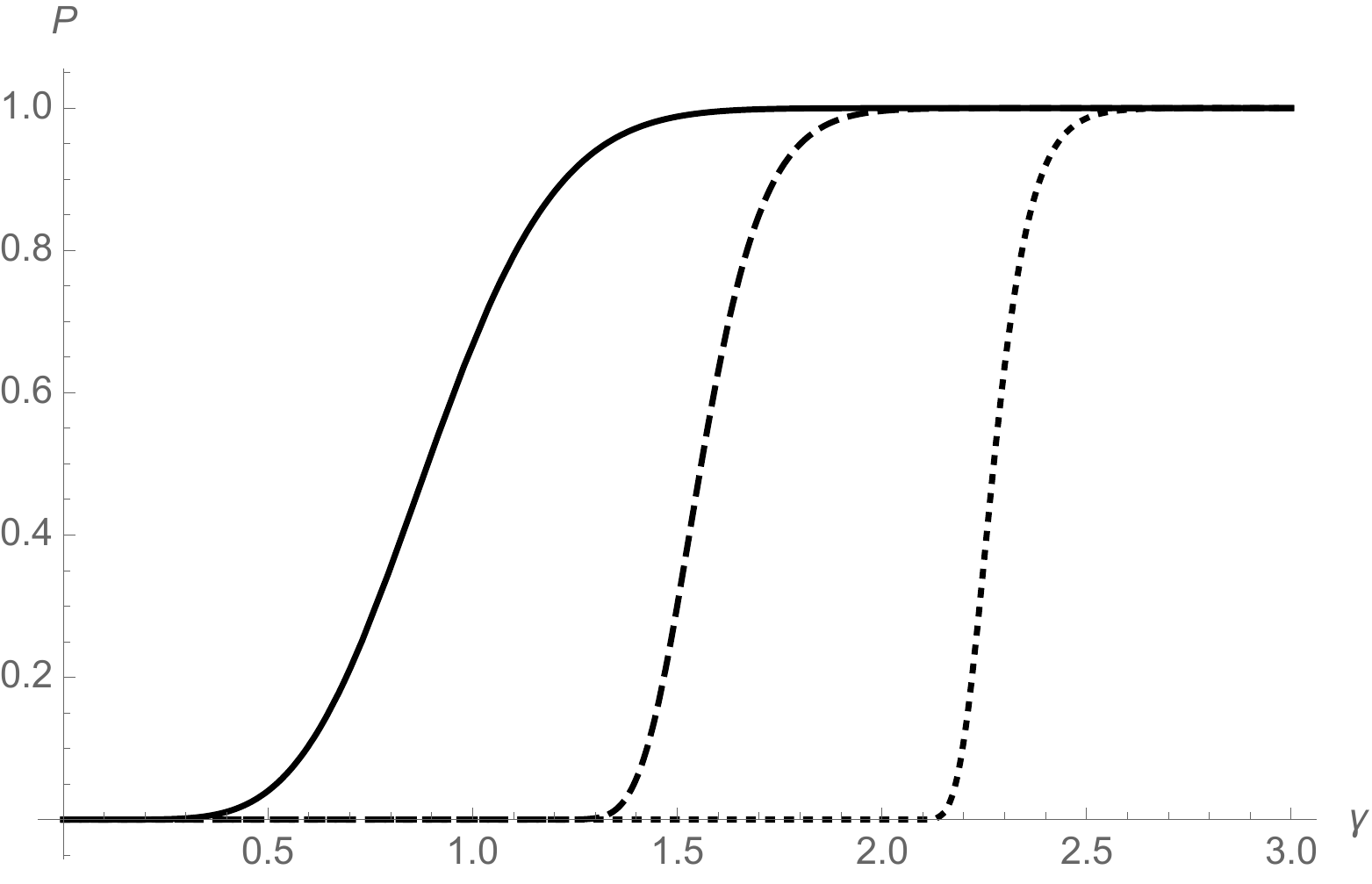}
\caption{Black hole probability~\eqref{1mpbh} as a function of $\gamma=\expec{\hat{R} ^{(+)} _{\rm H}}/2\,\lambda_\mu$
for $N=1$ (solid line), $N=10^2$ (dashed line) and $N=10^6$ (dotted line).}
\label{p<1}
\end{figure}
\begin{figure}[h]
\centering
\includegraphics[width=10cm]{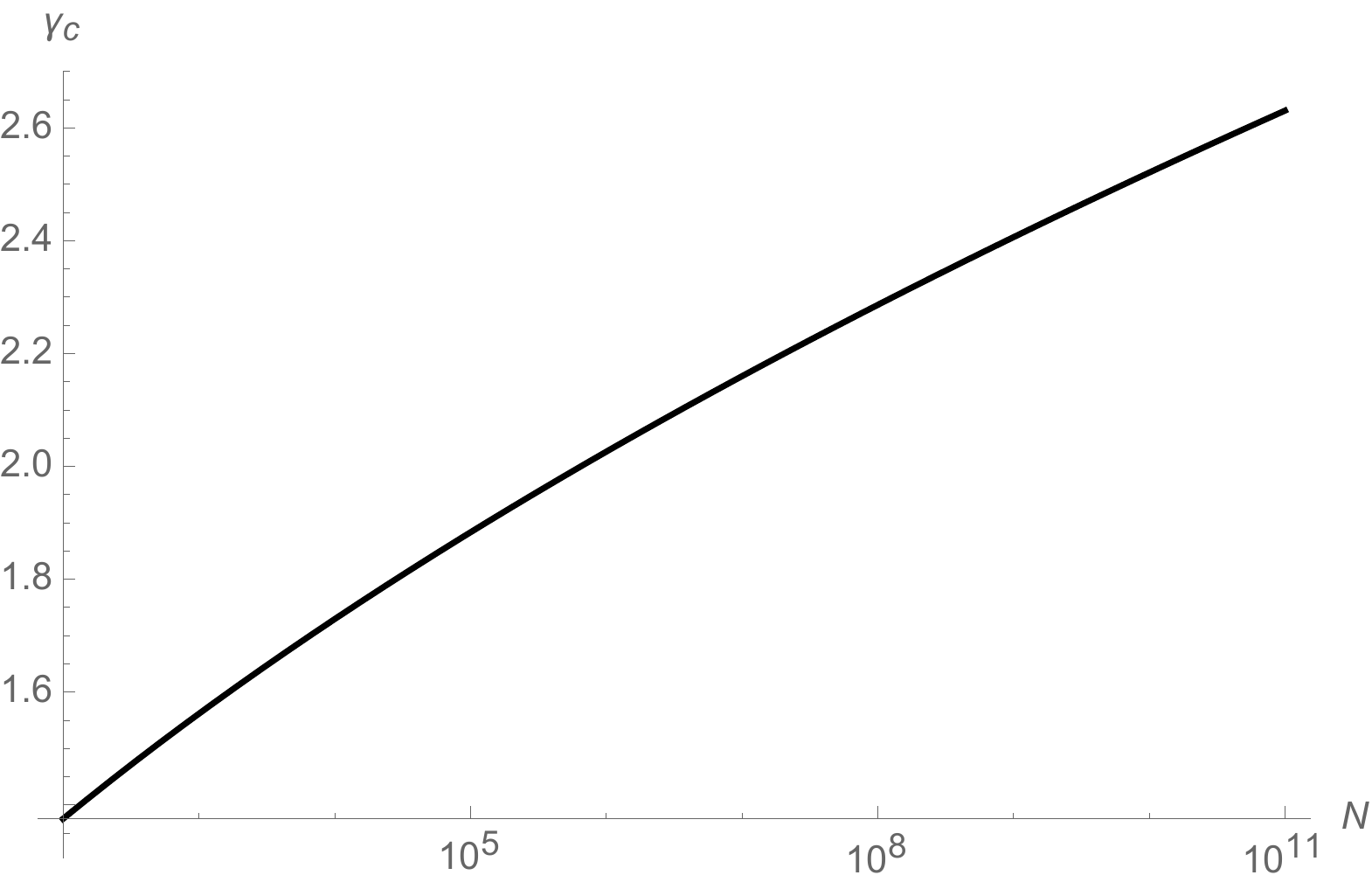}
\caption{Value of $\gamma_{\rm c}$ such that the black hole probability~\eqref{1mpbh} is given by
$P_{\rm BH}(n_+,N)=[P_{\rm BH}(\gamma\ge\gamma_{\rm c})]^N\ge 99\%$ for $N=10^2$ to $N=10^{11}$.}
\label{logG}
\end{figure}
\par
The single-particle ($N=1$) black hole probability $P_{(+)}(\gamma)$ is represented by the solid line
in Fig.~\ref{p<1}, from which it is clear that it practically saturates to 1 for $\gamma\gtrsim 2$. 
The same graph shows that the minimum value of $\gamma$ for which
$P_{\rm BH}(n_+,N)=[P_{(+)}(\gamma)]^N$ approaches 1 increases with $N$
(albeit very slowly).
For instance, if we define $\gamma_{\rm c}$ as the value at which $P_{\rm BH}(n_+,N)\simeq 0.99$, we
obtain the values of $\gamma_{\rm c}$ plotted in Fig.~\ref{logG}. 
It is also interesting to note that, for $\gamma=1$, which we saw can realise the extremal Kerr geometry,
we find 
\be
P_{\rm BH}(n_+,N)
\simeq
P_{\rm BH}(N)
\simeq
(0.67)^N
\ ,
\ee
and the system is most likely not a black hole for $N\gg 1$, in agreement with the probability density
shown in Fig.~\ref{Pr}.
One might indeed argue this probability is always too small for a (semi)classical black hole,
and that the extremal Kerr configuration is therefore more difficult to achieve. 
\par
Analogously, we can compute the probability $P_{\rm IH}$ that the inner horizon is realised.
Instead of Eq.~\eqref{ph1m}, we now have 
\be
{\mathcal{P}}_{(-)}(\Rh)
=
\delta(\Rh -\expec{\hat{R}^{(-)} _{\rm H}})
\ ,
\label{ph1mi}
\ee
which analogously leads to
\be
P_{\rm IH}(n_+,N)
=
\left[P_<^{(-)}(r<\expec{\hat R_{\rm H}^{(-)}})\right]^N
\ .
\ee
It is then fairly obvious that, for any fixed value of $\gamma$, $P_{\rm IH}(n_+,N)\le P_{\rm BH}(n_+,N)$ 
and that equality is reached at the extremal geometry with
$\expec{\hat R_{\rm H}^{(-)}}\simeq \expec{\hat R_{\rm H}^{(+)}}$.
Moreover, from $0\le L^2\le 1$ and Eq.~\eqref{1rh-},
we find $\expec{\hat R_{\rm H}^{(-)}}\lesssim \expec{\hat R_{\rm H}^{(+)}}/\gamma^2$,
so that for $\gamma=2$, the probability $P_{\rm IH}\lesssim (0.04)^N$ is totally negligible for $N\gg 1$.
This suggest that the inner horizon can remain extremely unlikely even in configurations that should represent
large (semi-)classical black holes.
\subsubsection{Superpositions}
The next step is investigating general superpositions of the states considered above,
\be
\ket{\Psi}
=
\sum_i
a_i\,\ket{M_i\, J_i}
\ ,
\ee
where $\sum_i|a_i|^2=1$ and 
\be
\ket{M_i \, J_i}
=
\bigotimes _{\alpha = 1} ^{N_i} \ket{g_i}_\alpha
\ ,
\ee
so that $M_i=N_i\,\mu_i$ and $J_i = (2\, N_{i+}\, -\, N_i)\, j_i\equiv N_i\, (2\, n_{i+}\, -\, 1)\, j_i$.
One can repeat the same analysis as the one performed for the single-mode
case, except that the two HWF's will now be superpositions of ADM values as well.
\par
In practice, this means that Eqs.~\eqref{ph1m} and \eqref{ph1mi} are now replaced by
\be
{\mathcal{P}}_{(\pm)}(\Rh)
=
\sum_i
|a_i|^2\,\delta(\Rh -{{R}^{(\pm)} _{{\rm H}_i}})
\ ,
\ee
where, from Eqs.~\eqref{hR+} and~\eqref{hR-}, the horizon radii are given by
\be
R_{{\rm H}_i}^{(\pm)}
=
\lp \frac{M_i}{\mpl}
\left(
1
\pm
\sqrt{1
-
\frac{\mpl^4\, J_i (J_i+1)}{M_i^4}}
\right)
\ ,
\label{expRhi}
\ee
and the expectation values of the horizon radii are correspondingly given by
\be
\expec{\hat R_{\rm H}^{(\pm)}}
=
\sum_i
|a_i|^2\, R_{{\rm H}_i}^{(\pm)}
\ .
\label{expRhab}
\ee
As usual, we obtain the probability that the system is a black hole by considering the outer horizon, for which
\be 
P_{\rm BH}
&\!\!=\!\!&
\sum_i
|a_i|^2\,
P_<^{(+)} (r_1 < {{R} ^{(+)} _{{\rm H}_i}},\ldots,r_{N_i} < {{R}^{(+)} _{{\rm H}_i}})
\nonumber
\\
&\!\!=\!\!&
\sum_i
|a_i|^2
\left[P_< ^{(+)} (r < {{R} ^{(+)} _{{\rm H}_i}})\right]^{N_i}
\ ,
\label{pbh2m}
\ee
where
\be
P_< ^{(+)} (r < {{R} ^{(+)} _{{\rm H}_i}})
=
\int_0^{{{R} ^{(+)}_{{\rm H}_i}}}
\mathcal{P}_{n_il_i}(r;\lambda_{\mu_i})\, \d r
\ ,
\ee
and
\be
\mathcal{P}_{nl}(r;\lambda_\mu)
=
\int_{-1}^{+1} \d \cos\theta
\int_0^{2\,\pi} \d\phi
\left| \psi_{nl\pm2}(r,\theta,\phi;\lambda_\mu) \right| ^2
\ . 
\ee
\par
The explicit calculation of the above probability immediately becomes very cumbersome.
For the purpose of exemplifying the kind of results one should expect,
let us just consider a state
\be
\ket{\Psi}
=
\frac{a\,\ket{M_1\, J_1}+b\,\ket{M_2\, J_2}}{\sqrt{|a|^2+|b|^2}}
\ ,
\ee
where the two modes in superposition are given by:
$N$ constituents with quantum numbers $n_1=0$, $l_1=2$ and $m=\pm 2$ in the state~\eqref{psi022},
here denoted with $\ket{g_1}$;
the same number $N$ of gravitons with quantum numbers $n_2=1$, $l_2=2$ and $m=\pm 2$ in the state
\be
\expec{\textbf{r} \, | \, g_2} 
= 
\psi_{12\pm2}(r,\theta,\phi)
=
\mathcal{N}_2 \,r^2\,e^{-\frac{r^2}{2\,\lambda_\mu^2}}\,\left(1-\frac{2\, r^2}{7\,\lambda_\mu^2} \right) \,Y_{2\pm2}(\theta,\phi)
\ ,
\label{psi122}
\ee
where we further assumed that all constituents have the same Compton/de~Broglie wavelength $\lambda_\mu$. 
It then follows that $M_1 = M_2 \equiv M$, so that
\be
{R_{{\rm H}_i}^{(\pm)}}
=
\lp \frac{M}{\mpl}
\left(
1
\pm
\sqrt{1
-
\frac{\mpl^4 \, J_i (J_i+1)}{M^4}}
\right)
\ ,
\label{expRhM}
\ee
and 
\be
\expec{\hat R_{\rm H}^{(\pm)}\!}\!
=\!
\lp \frac{M}{\mpl}\! \left[1 \pm \frac{1}{|a|^2\!+\! |b|^2}\!
\left(\! |a|^2 \sqrt{1-\frac{\mpl^4 \, J_1 (J_1+1)}{M^4}}\! +\! 
|b|^2\sqrt{1- \frac{\mpl^4 \, J_2 (J_2+1)}{M^4}} \right)\right]\!\! ,
\label{expRhM2}
\ee
with each of the $J_i$'s depending both on the numbers of spin up and the total number
of constituents of each type, as defined in the beginning of this section.
We also notice that when both $J_1$ and $J_2$ go to zero the expression simplifies to
\be
\expec{\hat R_{{\rm H}}^{(+)}}
=
\lp \frac{2\, M}{\mpl}
\ ,\label{expRhM3}
\ee
while $\expec{\hat R_{{\rm H}}^{(-)}}=0$, as expected for a Schwarzschild black hole. 
\par
\begin{figure}[t]
\centering
\includegraphics[width=16cm]{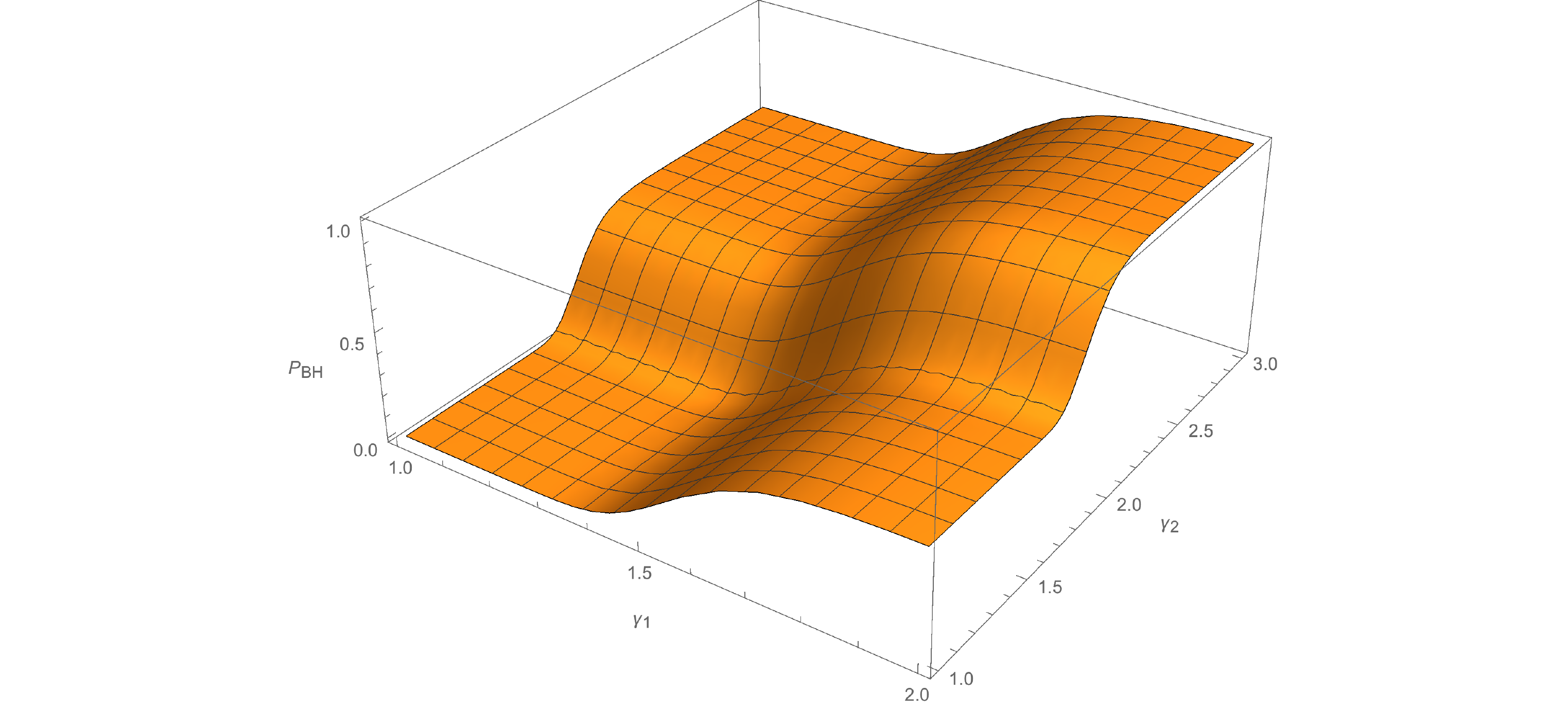}
\caption{Probability $P_{\rm BH}$ as a function of $\gamma_1={R_{\rm H_1}^{(+)}}/2\,\lambda_{\mu}$
and $\gamma_2={R_{\rm H_2}^{(+)}}/2\,\lambda_{\mu}$, for $a=b=1$ and $N=100$.}
\label{Pbh2modes100}
\end{figure}

The probability~\eqref{pbh2m} can be computed explicitly and is shown
in Fig.~\ref{Pbh2modes100} for $N=100$, with $a=b=1$. 
Beside the specific shape of those curves, the overall result appears in line with what we found in the
previous subsection for an Hamiltonian eigenstate: the system is most certainly a black hole provided
the Compton/de~Broglie length is sufficiently shorter than the possible outer horizon radius (that is,
for sufficiently large $\gamma_1$ and $\gamma_2$). 
\section{Conclusions}
\label{conc}
\setcounter{equation}{0}
After a brief review of the original HQM for static spherically symmetric sources,
we have generalised this formalism in order to provide a proper framework for the study of quantum
properties of the causal structure generated by rotating sources.
We remark once more that, unlike the spherically symmetric case~\cite{fuzzyh,hqft}, this extension is not based
on (quasi-)local quantities, but rather on the asymptotic mass and angular momentum of the Kerr class of
space-times.
As long as we have no access to local measurements on black hole space-times, this limitation
should not be too constraining.     
\par
In order to test the capabilities of the so extended HQM, one needs a specific (workable) quantum model
of rotating black holes.
For this purpose, we have considered the harmonic model for corpuscular black holes~\cite{QHBH}, which
is simple enough to allow for analytic investigations.
Working in this framework, we have been able to design specific configurations of harmonic black holes
with angular momentum and confirm that they are indeed black holes according to the HQM.
Some other results appeared, somewhat unexpected.
For instance, whereas it is reasonable that the probability of realising the inner horizon be smaller than 
the analogous probability for the outer horizon, it is intriguing that the former can indeed be negligible
for cases when the latter is close to one.
It is similarly intriguing that (macroscopic) extremal configurations do not seem very easy to achieve with
harmonic states.
\par
The results presented in this work are overall suggestive of interesting future developments and 
demand considering more realistic models for self-gravitating sources and black holes.
For example, it would be quite natural to apply the HQM to regular configurations of the kinds
reviewed in Refs.~\cite{nicolini,frolov,spallucci}. 
\section*{Acknowledgments}
R.C.~and A.G.~are partially supported by the INFN grant FLAG.
The work of A.G.~has also been carried out in the framework of the activities of the National Group
of Mathematical Physics (GNFM, INdAM).
O.M.~was supported by the grant LAPLAS~4. 
%

%

\begin{thebibliography}{99}
%
\bibitem{fuzzyh}
R.~Casadio,
``Localised particles and fuzzy horizons: A tool for probing Quantum Black Holes,''
arXiv:1305.3195 [gr-qc];
Springer Proc.\ Phys.\  {\bf 170} (2016) 225
[arXiv:1310.5452 [gr-qc]].
%
\bibitem{hqft}
R.~Casadio, A.~Giugno and A.~Giusti,
Gen.\ Rel.\ Grav.\  {\bf 49} (2017)  32
[arXiv:1605.06617 [gr-qc]].
%
\bibitem{gupf} 
R.~Casadio and F.~Scardigli,
Eur.\ Phys.\ J.\ C {\bf 74} (2014) 2685
[arXiv:1306.5298 [gr-qc]].
%
\bibitem{acmo}
R.~Casadio, O.~Micu and F.~Scardigli,
Phys.\ Lett.\ B {\bf 732} (2014) 105
[arXiv:1311.5698 [hep-th]].
%
\bibitem{Casadio:2015rwa} 
R.~Casadio, O.~Micu and D.~Stojkovic,
JHEP {\bf 1505} (2015) 096
[arXiv:1503.01888 [gr-qc]];
Phys.\ Lett.\ B {\bf 747} (2015) 68
[arXiv:1503.02858 [gr-qc]].
%
\bibitem{Casadio:2015qaq}
R.~Casadio, A.~Giugno and O.~Micu,
Int.\ J.\ Mod.\ Phys.\ D {\bf 25} (2016) 1630006
[arXiv:1512.04071 [hep-th]].
%
\bibitem{qlm-review}
L.~B.~Szabados,
Living Rev.\ Rel.\  {\bf 12} (2009) 4.
%
%
\bibitem{QHBH}
R.~Casadio and A.~Orlandi,
JHEP {\bf 1308} (2013) 025
[arXiv:1302.7138 [hep-th]].
%
\bibitem{muck}
W.~M\"uck and G.~Pozzo,
JHEP {\bf 1405} (2014) 128
[arXiv:1403.1422 [hep-th]].
%
\bibitem{dvali} 
G.~Dvali and C.~Gomez,
JCAP {\bf 01} (2014) 023
[arXiv:1312.4795 [hep-th]].
``Black Hole's Information Group'',
arXiv:1307.7630;
Eur.\ Phys.\ J.\ C {\bf 74} (2014) 2752
[arXiv:1207.4059 [hep-th]];
Phys.\ Lett.\ B {\bf 719} (2013) 419
[arXiv:1203.6575 [hep-th]];
Phys.\ Lett.\ B {\bf 716} (2012) 240
[arXiv:1203.3372 [hep-th]];
Fortsch.\ Phys.\  {\bf 61} (2013) 742
[arXiv:1112.3359 [hep-th]];
G.~Dvali, C.~Gomez and S.~Mukhanov,
``Black Hole Masses are Quantized,''
arXiv:1106.5894 [hep-ph];
R.~Casadio, A.~Giugno, O.~Micu and A.~Orlandi,
Entropy {\bf 17} (2015) 6893
[arXiv:1511.01279 [gr-qc]].
%
\bibitem{adm}
R.L.~Arnowitt, S.~Deser and C.W.~Misner,
Phys.\ Rev.\  {\bf 116} (1959) 1322.
%
\bibitem{Thorne:1972ji}
K.S.~Thorne,
``Nonspherical Gravitational Collapse: A Short Review,''
in {\em J.R.~Klauder, Magic Without Magic,\/} San Francisco (1972),
231.
%
\bibitem{baryons}
R.~Casadio, A.~Giugno and A.~Giusti,
Phys.\ Lett.\ B {\bf 763} (2016) 337
[arXiv:1606.04744 [hep-th]].
%
\bibitem{nicolini}
P.~Nicolini,
Int.\ J.\ Mod.\ Phys.\ A {\bf 24} (2009) 1229
[arXiv:0807.1939 [hep-th]].
%
\bibitem{frolov}
V.~P.~Frolov,
Phys.\ Rev.\ D {\bf 94} (2016) 104056
[arXiv:1609.01758 [gr-qc]].
%
\bibitem{spallucci} 
E.~Spallucci and A.~Smailagic,
``Regular black holes from semi-classical down to Planckian size,''
arXiv:1701.04592 [hep-th].
%
\end{thebibliography}
\end{document}